\documentstyle[aps,pra,epsf,floats]{revtex}

\begin{document}

\twocolumn[\hsize\textwidth\columnwidth\hsize\csname@twocolumnfalse%
\endcsname

\title{Critical exponents of random XX and XY chains:\\
       Exact results via random walks}
 
\author{Heiko Rieger$^{1,2,3}$, R\'obert Juh\'asz$^{4}$ 
        and Ferenc Igl\'oi$^{3,4,5}$}
\address{
   $^1$ NIC c/o Forschungszentrum J\"ulich, 52425 J\"ulich, Germany\\
   $^2$ Institut f\"ur Theoretische Physik, Universit\"at zu K\"oln, 
   50937 K\"oln, Germany\\
   $^3$ Laboratoire de Physique des Mat\'eriaux, Universit\'e Henri
        Poincar\'e, BP 239, F-54506 Vand{\oe}vre l\'es Nancy Cedex, France\\
   $^4$ Institute for Theoretical Physics,
        Szeged University, H-6720 Szeged, Hungary\\
   $^5$ Research Institute for Solid State Physics and Optics,   
        H-1525 Budapest, P.O.Box 49, Hungary}

\date{June 14, 1999}

\maketitle

\begin{abstract}  
  We study random XY and (dimerized) XX spin-$1/2$ quantum spin chains
  at their quantum phase transition driven by the anisotropy and
  dimerization, respectively. Using exact expressions for
  magnetization, correlation functions and energy gap, obtained by the
  free fermion technique, the critical and off-critical
  (Griffiths-McCoy) singularities are related to persistence
  properties of random walks. In this way we determine exactly the
  decay exponents for surface and bulk transverse and longitudinal
  correlations, correlation length exponent and dynamical exponent.
\end{abstract}



]

\newcommand{\bc}{\begin{center}}
\newcommand{\ec}{\end{center}}
\newcommand{\be}{\begin{equation}}
\newcommand{\ee}{\end{equation}}
\newcommand{\beqn}{\begin{eqnarray}}
\newcommand{\eeqn}{\end{eqnarray}}
\newcommand{\ba}{\begin{array}}
\newcommand{\ea}{\end{array}}

\narrowtext

Disordered quantum spin chains have gained much interest recently
\cite{fisher,fisherxx,rtic,kunyang,mckenzie,fabrizio,monthus,sigrist}.
It seems to be established right now that the critical properties in
these one-dimensional system are governed by an infinite-disorder
fixed-point \cite{fisherb} and the application of a renormalization
group (RG) scheme \'a la Dasgupta and Ma \cite{dasguptama} is a powerful
tool to determine critical properties and static correlations of these
new universality classes, either analytically, if possible, or
numerically. Although the underlying renormalization scheme is
extremely simple the analytical computations are sometimes tedious
\cite{fisher,fisherxx}. Therefore an alternative route to the exact
determination of critical exponents and other quantities of interest
is highly desirable, and this is what we are going to present in this
letter. In doing so we follow a route on which we already traveled
successfully for the random transverse Ising chain
\cite{big1d,diffusion,avpers}, and here we are going to do one step
further studying random XX and XY models with the help of a
straightforward and efficient mapping to random walk problems. This
mapping is not only a short-cut to the results known from analytical
RG calculations, it also gives new exact results in the off-critical
region (the Griffiths-phase \cite{griffiths}) and provides a mean to study
situations in which the RG procedure must fail, as for instance in the
case of correlated disorder \cite{corrdis}. Here we confine ourselves
to a concise presentation of the basic ideas including the
determination of various exponents for the first time. The technical
details of the derivations and further results are deferred to a
subsequent publication \cite{big}.

The model that we consider is a spin-$1/2$ XY--quantum spin chain with
$L$ sites and open boundaries, defined by the Hamiltonian
\be
H=\sum_{l=1}^{L-1}\left( J_l^x S_l^x S_{l+1}^x+ J_l^y S_l^y S_{l+1}^y
\right)\;,
\label{hamil}
\ee
where the $S_l^{x,y}$ are spin-$1/2$ operators and the interaction
strengths or couplings $J_l^{x,y}>0$ are independent random variables
modeling quenched disorder. In the case of the random XY chain one
has two independent distributions for the couplings $J^x$ and $J^y$,
$\rho^x$ and $\rho^y$, respectively, whereas the random dimerized
XX-chain has perfectly isotropic couplings $J_l^x=J_l^y=J_l$ but two
independent probability distributions for the even and odd couplings
(i.e.\ for $J_{2l}=J_{2l}^e$ and $J_{2l-1}=J_{2l-1}^o$), $\rho^e$ and $\rho^o$,
respectively.

The model (\ref{hamil}) has a critical point given by $[\ln J^x]_{\rm
  av} = [\ln J^y]_{\rm av}$ in the XY case and $[\ln J^e]_{\rm
  av}=[\ln J^o]_{\rm av}$ in the XX case (here $[\ldots]_{\rm av}$
denotes the disorder average). The distance from the critical point is
  conveniently measured in the variable
\be
\delta={[\ln J^{x(e)}]_{\rm av} - [\ln J^{y(o)}]_{\rm av}\over
       {\rm var}[\ln J^{x(e)}]+{\rm var}[\ln J^{y(o)}]}\;,
\label{aniz}
\ee
where ${\rm var}(x)$ is the variance of random variable $x$. At the
critical point ($\delta=0$) spatial correlations decay
algebraically, for instance in a finite system of length $L$ with
{\it periodic} boundary conditions the bulk-correlations decay as
\be
[C^{\mu}(L)]_{\rm av}
=[\langle0|S^{\mu}_1S^{\mu}_{L/2}|0\rangle]_{\rm av}
\sim L^{-\eta^{\mu}}
\label{corr}
\ee
for $\mu=x,y,z$, $\langle0|$ denotes the ground state of
(\ref{hamil}), whereas for a finite system of length $L$ with {\it open}
boundary conditions the end-to-end correlations decay with a different
exponent like
\be
[C_1^{\mu}(L)]_{\rm av}
=[\langle0|S^{\mu}_1S^{\mu}_{L}|0\rangle]_{\rm av}
\sim L^{-\eta_1^{\mu}}\;.
\label{endcorr}
\ee
Away from the critical point ($\delta\ne0$) the infinite system
develops long range order. For the $XY$ model $\lim_{L\to\infty}[C^{\mu}(L)]_{\rm av}
=(m^\mu)^2\ne0$, with $m^x>0$ for $\delta>0$ and $m^y>0$ for
$\delta<0$, whereas for the $XX$ model there is dimerization for $\delta \ne 0$
with non-vanishing string order\cite{girvin}. One can introduce local transverse and longitudinal order
parameters $m_l^{x,y}$ and $m_l^z$ also for a finite system (with
open boundaries) using the off-diagonal matrix element
$[m_l^\mu]_{\rm av} = [\langle1|S^{\mu}_{1}|0\rangle]_{\rm av}$, where
$\langle1|$ is the lowest excited state with a non-vanishing matrix-element
\cite{turbanigloi}. Analogous to bulk and
end-to-end correlations the bulk and surface magnetizations $m_{L/2}^\mu$
and $m_1^\mu$, respectively, behave differently:
\be
[m_{L/2}^\mu]_{\rm av}\sim L^{-x^\mu}
\quad{\rm and}\quad
[m_1^\mu]_{\rm av}\sim L^{-x_1^\mu}
\quad(\delta=0)
\ee
where the critical exponents $x^\mu$ and $x_1^\mu$ fulfill
the scaling relation $2x^\mu=\eta^\mu$ and $2x_1^\mu=\eta_1^\mu$.

Now we are going to determine the critical exponents introduced above.
We use the free-fermion representation of model (\ref{hamil}) to
derive exact expressions for the local order-parameters whose finite
size scaling behavior follows then from a mapping to a random
walk problem. 

We start with the longitudinal order parameter for the random XX chain
(of length $L$ with open boundaries; for convenience we assume from
now on that $L$ is a multiple of $4$), which is given by \cite{big}
%
\beqn
m_{2l-1}^z(XX)=
\frac{1}{2}\Biggl\{1
&+&\displaystyle
\sum_{k=l}^{L/2-1}\prod_{j=l}^k\biggl(\frac{J_{2j-1}}{J_{2j}}\biggr)^2
\nonumber\\
&+&
\displaystyle
\sum_{k=1}^{l-1}\prod_{j=1}^k\biggl(\frac{J_{2l-2j}}{J_{2l-2j-1}}\biggr)^2
\Biggr\}^{-1}
\label{magz}
\eeqn
%
for odd sites. Note that the couplings to the left and to the right of
the spin that is considered enter this expression differently. For the
surface ($l=1$) , where one only has "right" couplings, this expression
is similar to an analogous result for the random transverse Ising
chain \cite{big1d} and its scaling properties are related to the survival
probability of a random walk with $L/2$ steps. This is easy to see for
the extreme binary distribution, in which $J_{2j}=1$ and
$J_{2j-1}=\lambda,\;\lambda^{-1}$ with probability $1/2$, taking the
limit $\lambda\to0$ (i.e.\ $\lambda^{-1}\to\infty$). Due to the
occurance of infinite terms in the sum in the denominator of the
r.h.s.\ of (\ref{magz}) one can easily identify those instances that
give a non-vanishing surface magnetization: When
$\forall k=1,\ldots,L/2-1: \prod_{j=1}^k J_{2j-1}<\infty$
the expression on the r.h.s. of (\ref{magz}) attains a non-vanishing
value (typically 1 or, less frequently, some fraction $1/n$),
otherwise it is zero. One can represent the disorder configuration
$J_1, J_3, J_5,\ldots,J_{L-1}$ as one instance of a random walk with
$L/2-1$ steps by saying that the walker in the $i$-th steps moves
downwards if $J_{2i-1}=\lambda$ and upwards if
$J_{2i-1}=\lambda^{-1}$, as it is sketched in Fig.1. In this way the
disorder configuration with non-vanishing surface magnetization
$m_1^z$ are easily identified: they represent surviving walks, i.e.\ 
walks that never move into the upper half. Thus $[m_1^z(XX)]_{\rm av}$
scales like the survival probability $P_{\rm surv}(L/2)$ of a random
walk with $L/2$ steps that vanishes like $L^{-1/2}$, i.e.
$[m_1^z(XX)]_{\rm av}\sim L^{-1/2}$. Therefore
\be
x_1^z(XX)=\frac{1}{2}\quad{\rm and}\quad\eta_1^z(XX)=1\;.
\label{surfxxz}
\ee
Inspecting the expression (\ref{magz}) for the bulk ($l=L/4$) one sees
that now (again for the extreme binary distribution) a
nonvanishing magnetization $m_{L/2-1}^z$ arises only if
$\forall k=L/4,\ldots,L/2-1: \prod_{j=L/4}^k J_{2j-1}<\infty$
and
$\forall k=1,\ldots,L/4-1: \prod_{j=1}^k J_{L/2-2j-1}^{-1}<\infty$.
We represent the disorder configuration to the right of the central
site, $J_{L/2-1},J_{L/2+1},\ldots,J_{L-1}$, as a random walk with
$L/4$ steps in the way as for the surface spin. The disorder
configuration $J_{L/2-1},J_{L/2-3},\ldots,J_1$ to the left is
represented as a second (independent) random walk also with $L/4$
steps, now counting backwards and with the step-direction reversed
(i.e.\ downwards for $J=\lambda^{-1}$ and upwards for $J=\lambda$),
since now strong bonds on odd sites imply weak coupling of the central
spin.  For illustration this representation is depicted in Fig. 1.
Now, for the bulk magnetization $m_{L/2-1}^z$ to be non-vanishing,
{\it both} halfs of the coupling configuration have to represent
surviving random walks. Thus the probability for a non-vanishing
magnetization $m_{L/2-1}^z$ is just the product of two survival
probabilities (since both walks are independent), i.e.\ 
%
$[m_{L/2-1}^z]_{\rm av}\sim\{P_{\rm surv}(L/4)\}^2\sim L^{-1}$
%
and therefore
\be
x^z(XX)=1\quad{\rm and}\quad\eta^z(XX)=2\;.
\ee

\begin{figure}
\epsfxsize=\columnwidth\epsfbox{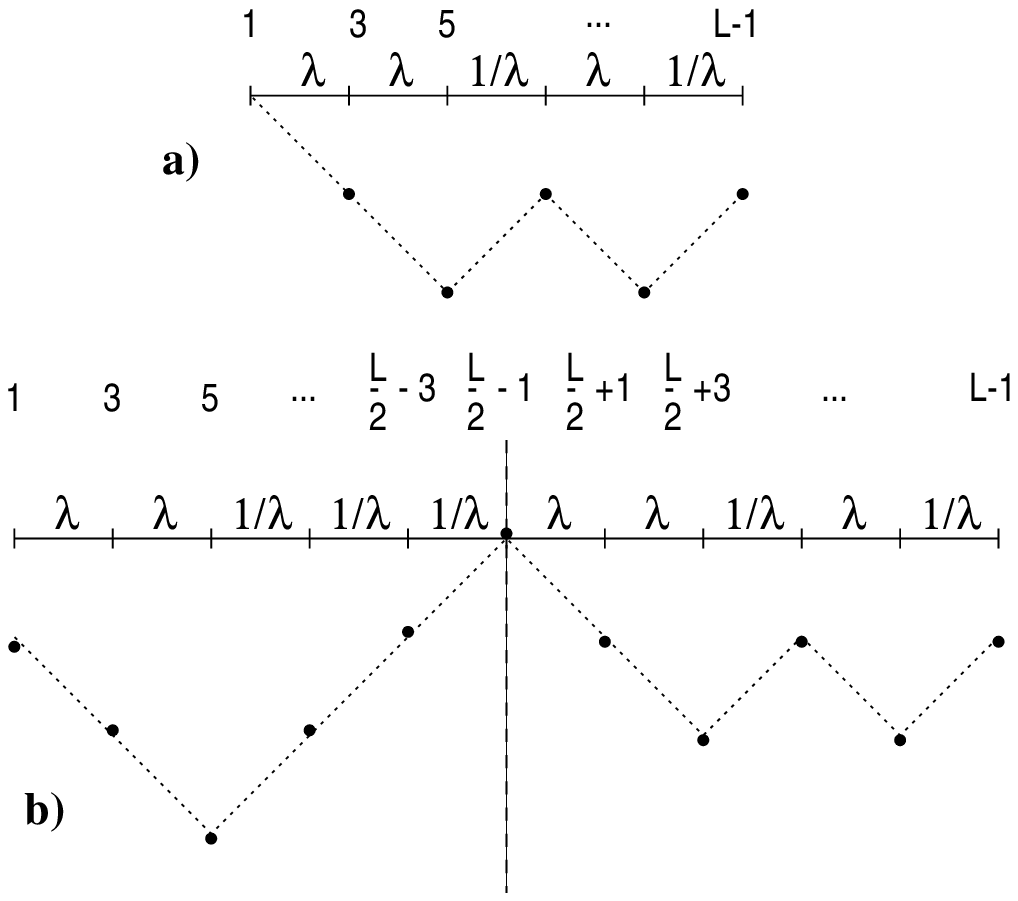}
\caption{Sketch of the configuration of odd bonds for a chain of length
  $L$ that gives a non-vanishing longitudinal magnetization
  $m_i^z\sim{\cal O}(1)$ for the surface spin, $i=1$, in (a) and the
  central spin, $i=L/2-1$, in (b). The example is for the extreme
  binary distribution with $J_{2i}=1$. Weak couplings
  ($J_{2i-1}=\lambda$) correspond to downward steps of the random
  walk, strong couplings ($J_{2i-1}=\lambda^{-1}$) to upwards steps.
  The walk in (a) has surviving character, it does not enter the upper
  half plane. In (b) one can identify two random walks each starting
  at the central site, $i=L/2-1$, one to the right and one to the
  left, and each of them has surviving character.}
\label{fig1}
\end{figure}

For the XY chain one has the following exact relation \cite{big} for
the disorder averaged longitudinal magnetization
\be
[m_l^z(XY)]_{\rm av}=[\{m_l^z(XX)\}^{1/2}]_{\rm av}^2\;,
\label{xyxxrel}
\ee
which yields immediately the surface and bulk exponents for
longitudinal order parameter and correlations: Since
$\{m_1^z(XX)\}^{1/2}$ has a non-vanishing value if and only if 
$m_1^z(XX)$ is non-vanishing, one obtains $[m_1^z(XY)]_{\rm av}\sim
\{P_{\rm surv}(L/4)\}^2\sim L^{-1}$, which means
\be
x_1^z(XY)=1\quad{\rm and}\quad\eta_1^z(XY)=2\;,
\label{surfxyz}
\ee
and further, $[m_{L/2-1}^z(XY)]_{\rm av}\sim
\{P_{\rm surv}(L/4)\}^4\sim L^{-2}$, which means
\be
x^z(XY)=2\quad{\rm and}\quad\eta^z(XY)=4\;.
\ee

For the transverse surface order parameter $m_1^x$ one has \cite{big}
the exact formula (valid for the XX and XY case)
\be
m_1^x=
{1 \over 2} \left[1+\sum_{l=1}^{L/2-1} \prod_{j=1}^l
\left( J^{y(o)}_{2j-1} \over J^{x(e)}_{2j} \right)^2\right]^{-1/2}\;,
\ee
which is similar to (\ref{magz}) for $m_1^z(XX)$, apart from the power
$1/2$ on the r.h.s., which again is only non-vanishing for
disorder configurations that represent surviving walks \'a la Fig.1a.
This then yields without any effort
\be
x_1^x(XX,XY)=1\quad{\rm and}\quad\eta_1^x(XX,XY)=2\;,\\
\label{surfxxx}
\ee

For the transverse bulk order parameter in the XY chain we use the
fact that the model can be mapped onto two transverse Ising models
(TIM), with uncorrelated disorder in both chains \cite{fisherxx,big}.
Through this mapping one obtains for the transverse correlation
function $C_{2i,2i+2r}^x=\langle0|S_{2i}^x S_{2i+2r}^x|0\rangle$ the
following identity \cite{big}
\beqn
[C_{2i,2i+2r}^x(XY)]_{\rm av} = 
4 && [C_{i,i+r}^x(TIM_{\rm free})]_{\rm av}
\label{corrxy}\\
\cdot& & [C_{i,i+r}^x(TIM_{\rm fixed})]_{\rm av}
\sim r^{-2\eta^x(TIM)}\;, \nonumber
\eeqn
where {\it fixed} and {\it free} indicated the boundary conditions.
Since the correlation function exponent is known exactly
\cite{fisher} to be $\eta^x(TIM)=(3-\sqrt{5})/2$ we have:
\be
x^x(XY)=(3-\sqrt{5})/2\quad{\rm and}\quad\eta^x(XY)=3-\sqrt{5}\;.
\ee
For the XX chain the two transverse Ising chains have perfectly
correlated disorder, which implies that the disorder averaged
transverse correlations do not factorize into two independent averages
as in (\ref{corrxy}). Therefore, for the transverse order parameter
exponent in the XX case we have to use a different route: The first
important observation is that the transverse bulk order parameter
$m_{L/2}^x=\langle1|S_{L/2}^x|0\rangle$ attains its maximum value
$1/2$ if the central spin is decoupled from the rest of the system,
i.e.\ when $J_{L/2-1}=J_{L/2}=0$. More generally we expect that
$m_{L/2}^x\sim{\cal O}(1)$ when it is weakly coupled to the rest of
the system. "Weakly coupled" in the case of the extreme binary
distribution means that the bond configuration to the left and to the
right of the central spin represent both surviving random walks, as
exemplified in Fig.1b (this is actually equivalent to the (exact)
condition for the longitudinal order parameter $m_{L/2}^z(XX)$ to be
non-vanishing).  This correspondence implies
%
$[m_{L/2}^x(XX)]_{\rm av}\sim\{P_{\rm surv}(L/4)\}^2\sim L^{-1}$
%
from which one obtains
\be
x^x(XX)=1\quad{\rm and}\quad\eta^x(XX)=2\;.
\label{etaxxx}
\ee
We verified the strong correlation between weak coupling and
non-vanishing transverse order parameter numerically in the following
way: We considered a chain with $L+1$ sites and the couplings at both
sides of the central spin were taken randomly in the form of surviving
walk character, where we used the binary distribution with
$\lambda=0.1$. For such small value of $\lambda$ the surface
order-parameter averaged over the surviving walk (sw) configurations
$[m_1^x]_{\rm sw}$ was very close to the maximal value of $1/2$. Then
we calculated numerically the order-parameter at the central spin and
its average value over surviving walk configurations $[m_{L/2}^x]_{\rm
  sw}$ as given in Table I.

\bc
\begin{tabular}{|c|c|c|}
\hline
L&$2[m_1^x]_{\rm sw}$&$2[m_{L/2}^x]_{\rm sw}$\\
\hline
32&0.994&0.764\\
64&0.991&0.682\\
128&0.991&0.647\\
256&0.991&0.577\\
\hline
\end{tabular}
\ec
{\small TABLE I: Surface and bulk transverse order-parameters averaged 
over 50000 surviving walk coupling configurations for the binary distribution
($\lambda=0.1$).}
\bigskip

As seen in the Table the averaged surface order-parameter stays
constant for large values of $L$, whereas the bulk order-parameter
decreases very slowly, actually slower than any power. The data can be
nicely fitted by $[m_{L/2}^x]_{\rm sw} \sim (\ln L)^{-1/2}$. Thus we
conclude that the numerical results confirm 
the exponents given in (\ref{etaxxx}), however there are
strong logarithmic corrections, 
which imply for the average transverse correlations
\be
[C^x(r)]_{\rm av} \sim r^{-2} \ln^{-1}(r)~~~{\rm XX-model}\;.
\label{CxXX}
\ee
These strong logarithmic corrections render the numerical calculation
of critical exponents very difficult \cite{girvin,stolzexx}. In
earlier numerical work using smaller finite systems disorder dependent
exponents were reported\cite{stolzexx}.  We believe that these
numerical results can be interpreted as effective, size-dependent
exponents and the asymptotic critical behavior is indeed described by
Eq. (\ref{CxXX}).

Away from the critical point the correlation length exponent $\nu$ can
be determined by the scaling behavior of the longitudinal surface
magnetization, i.e.\ (\ref{magz}) with $l=1$, which can be inferred
from the survival properties of a, now biased, random walk:
%
$[m_1^{x,y}(\delta,L)]_{\rm av} \sim P_{\rm surv}(\delta,L/2)$.
%
A non-vanishing distance $\delta$, see (\ref{aniz}), from the critical
point means that the disorder configurations can be represented by
random walk that have a drift either towards ($\delta>0$) or away from
($\delta<0$) an absorbing wall (take for instance the extreme binary
distribution, in which weak bonds $\lambda$ occur with a probability
$(1-\delta)/2$ and strong bonds $\lambda^{-1}$ occur with a
probability $(1+\delta)/2$ and compare with Fig.1a). Recalling the
asymptotic properties of the survival probability of random walks
\cite{big1d} one gets for $\delta>0$
$P_{\rm surv}(\delta>0,L ) \sim \exp(-L/\xi)$ with $\xi \sim
\delta^{-2}$, where the characteristic length scale $\xi$ of surviving
walks corresponds to the {\it average} correlation
length of the XX and XY chains:
\be
[\xi]_{\rm av}\sim\delta^{-\nu}\quad{\rm with}\quad\nu=2\;.
\label{nuav}
\ee
For $\delta<0$ the drift away from the adsorbing wall yields a finite 
survival probability even in the infinite system, $P_{\rm
  surv}(\delta<0,L/2)\propto\delta$, which implies that 
$[m_1^\mu]_{\rm av}\propto|\delta|^{-\beta_1^\mu}$, with
  $\beta_1^{x,y}(XX,XY)=\beta_1^z(XX)=1$ and $\beta_1^z(XY)=2$.
Since $x_1^\mu=\beta_1^\mu/\nu$ and $\nu=2$ from (\ref{nuav}) one
reconfirms the results about the surface magnetization exponents 
in (\ref{surfxxz}), (\ref{surfxyz}) and (\ref{surfxxx})

The typical correlation length, $\xi_{\rm typ}$ can be inferred from
the scaling behavior of the typical surface magnetization $\ln m_1\sim
\sum_j\{\ln(J_{2j-1}^{y(o)})-\ln(J_{2j}^{x(e)})\}\propto \delta L$,
which gives
\be
[\xi]_{\rm typ}\sim\delta^{-\nu_{\rm typ}}
\quad{\rm with}\quad\nu_{\rm typ}=1
\ee

The critical and off-critical scaling behavior of the low energy
exciations and dynamical correlations can be deduced from the formula
for the gap $\epsilon$ \cite{big}
\be
\epsilon(L) = m_1^x m_{L-1}^x J^y_{L-1}\prod_{j=1}^{L/2-1} 
{J^{y(o)}_{2j-1} \over J^{x(e)}_{2j}} \;,
\label{gap}
\ee
which is analogous to a corresponding formula for the gap in the
transverse Ising chain \cite{big1d}. At the critical point one
observes that $\ln\epsilon$ is a sum of $L$ independently distributed
random variables with zero mean (since $\delta=0$), for which the
central limit theorem applies. Therefore the probability distribution
of gaps obeys 
%
$P(\ln\epsilon)\sim L^{-1/2}\tilde{p}(\ln\epsilon/L^{-1/2})$
%
and one uses scaling arguments as in \cite{dynamics} to deduce the
asymptotic (imaginary) time dependence of the spin-spin
autocorrelation function
$G_l^\mu(\tau)=[\langle0|S_l^\mu(\tau)S_l^\mu(0)|0\rangle]_{\rm av}$.
\be
G_a^\mu(\tau)\sim(\ln\tau)^{-\eta_a^\mu}
\ee
for the surface ($a=1$) and bulk ($a=$bulk), respectively, with the
critical exponents $\eta_a^\mu$ as given above. 

Away from the critical point in the Griffiths-phase
\cite{griffiths} the gap distribution has still an algebraic
tail
%
$P(\epsilon) \sim \epsilon^{-1+1/z'(\delta)}\;,$
%
with a dynamical exponent $z'(\delta)$ that varies continuously with
the distance from the critical point $\delta$ and is given by the
exact (implicite) formula \cite{diffusion}
\be
\left[\left( J^{x(e)} \over J^{y(o)} \right)^{1/z'(\delta)} 
\right]_{\rm av}=1\;.
\label{zexact}
\ee
The dynamical exponent $z'(\delta)$ parameterizes all Griffiths-McCoy
singularities occurring in the Griffiths-phase, e.g.\ 
the spin-spin autocorrelations decay algebraically as
\be
G_l(\tau)\sim\tau^{-1/z'(\delta)}\;,
\ee
which gives for the susceptibility $\chi^\mu\sim T^{-1+1/z'(\delta)}$
diverging for $T\to0$ ($T$ = temperature) if $z'(\delta)>1$.

To summarize we have shown how to obtain a complete description of the
critical and off-critical singularities of random XX and XY chains
with simple random walk arguments using exact formulas arising from
the free fermion description of these quantum spin models. All results
for the critical exponents are therefore exact. One should note that
for the transverse bulk order parameter exponent for the XY model we
referred to a result for the transverse Ising model obtained by a RG
calculation \cite{fisher} and for the same exponent of the XX model we
showed the existence of strong logarithmic corrections.

F.\ I.'s work has been supported by the Hungarian National Research
Fund under grant No OTKA TO23642, TO25139, MO28418 and by the
Ministery of Education under grant No. FKFP 0596/1999.  H.\ R. was
supported by the Deutsche Forschungsgemeinschaft (DFG). The
Laboratoire de Physique des Mat\'eriaux is Unit\'e Mixte de Recherche
CNRS No 7556.

\end{document}